# Laser-scanning of induction-melted Al alloys: are they representative of additively manufactured ones?

Authors: Zhaoxuan Ge[1], Sebastian Calderon[1], S. Mohadeseh Taheri-Mousavi[1,2*]

[1]Department of Materials Science and Engineering, Carnegie Mellon University, 5000 Forbes Avenue, Pittsburgh, PA 15213, USA.

[2]Department of Mechanical Engineering, Carnegie Mellon University, 5000 Forbes Avenue, Pittsburgh, PA 15213, USA.

*Corresponding author: S. Mohadeseh Taheri-Mousavi, smtaherimousavi@cmu.edu

**Highlights:**

- Compared microstructures and hardness of laser-scanned cast samples with LPBF ones.
- Microstructure is similar at high magnification, with grain/phase size difference under 20%.
- Hardness of the laser-scanned samples can relatively represent that of LPBF ones.

**Abstract:**

The bottleneck of alloy design for powder-based additive manufacturing (AM) resides in customized powder production—an expensive and time-consuming process hindering the rapid closed-loop design iterations. This study analyzed an expedited experimental workflow, i.e., multipath laser scanning of induction-melted samples, to mimic rapid solidification of AM to serve as an alternative approach to down-select from the design space. Using Al-Ni-Zr-Er model alloy, comprehensive multi-scale characterizations were performed to compare microstructural features between laser-scanned and laser powder bed fusion (LPBF) samples. Although demonstrating a difference in melt pool geometries, the microstructures in scanning electron microscopy (SEM)- and transmission electron microscopy (TEM)- scale demonstrate a high degree of similarity, in terms of microstructure morphology, grain size, presence of precipitates, and phase distribution. The mechanical performance was evaluated by microhardness tests. The results revealed a 20% reduction in laser-scanned samples compared to LPBF samples, attributed to the thermal history and potential differences in phase fractions. The decreasing trend was also observed in the benchmark alloy showing a 10% absolute error with respect to the model alloy. This study underscores the potential of this workflow to accelerate alloy design in AM by circumventing customized powder production and encourages further exploration across diverse materials and processing parameters.

## Traditional AM alloy design workflow

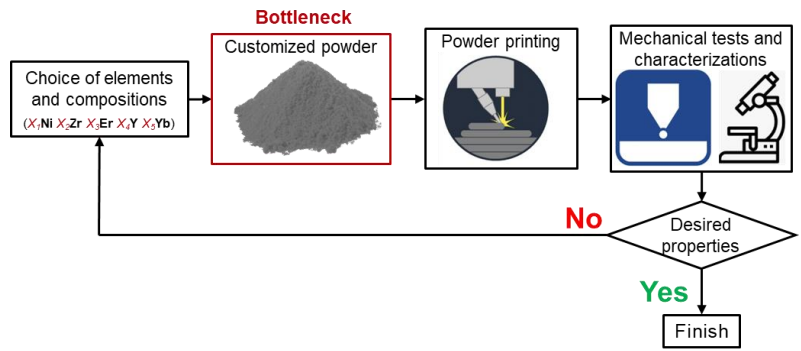

## Rapid experiment workflow for AM alloy design workflow

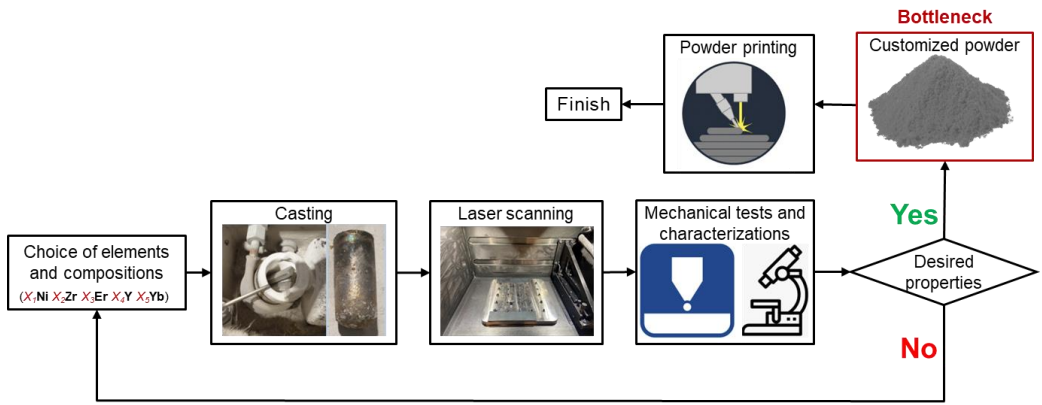

# 1. Introduction:

AM has major advantages of shortening supply chains for market distribution and realization of intricate geometric components. However, our current understanding of the underlying physics of AM is still limited due to the complicated thermal history and various heat transfers. Thus, the community still predominantly focuses on replicating reliable properties of cast/wrought alloys for AM parts. Despite these limitations, the inherent characteristics of AM pave a new pathway for alloy design. For example, the layer-wise fabrication of AM exhibits the capability to circumvent the occurrence of macrosegregation when incorporating heavy elements into the material matrix. Furthermore, LPBF as one of AM techniques has $10^5$-$10^7$ K/s high cooling rates [1], which encompasses advantages including supersaturation of solutes [2], [3], [4], refinements of microstructural characteristics [5], [6], [7], [8], [9], augmentation in the volume fraction and/or number densities of strengthening phases [10], [11], and the ability to harness phase metastability [11], [12], [13]. These attributes offer a realm of design flexibility, fostering opportunities for innovative alloys that can unleash the economic potential of AM beyond the realm of prototyping.

To design a new alloy with AM, the process starts with the choice of the alloying elements and their concentrations. This choice can be made by performing e.g., calculation of phase diagram (CALPHAD)-based simulations. Upon identification of a specific composition, we make the customized powder and print them. Then, an array of characterization and performance evaluations are conducted to validate desired properties. Upon successful attainment of the target properties, the alloy design process concludes. However, the design usually requires several iterations to discover the optimal composition because of the difficulty in satisfying multiple physical constraints that can be mutually exclusive. There are also uncertainties between numerical predictions and experimental validations. Among all the steps in the design loop, the critical bottleneck is in the production of custom powders [14]. This entails several steps, beginning with the ordering of ingots from vendors and awaiting their delivery. For those lack of cast and powder manufacturing resources, the ingots must be ordered from vendors and then sent to another company for powder production. Once the custom powder is fabricated, a protracted process ensues for receiving and safely storing the powder given the flammable and explosive characteristics of powders. These procedures not only incur significant costs but also entail an extremely long lead time.

Considering the limited availability of the customized powder, high-throughput experiments such as the direct printing of chemical gradients within the specimen have been proposed to expedite the alloy design in the AM field [15], [16], [17], [18]. However, gradient printing is only applicable to certain technologies such as direct energy deposition (DED) [19], [20], [21] and ink jetting [22]. Even though the compositional gradient printing technique aims to generate integrated specimens that encapsulate substantial information from limited samples, powder production is still a necessity. In response to this challenge, an expedited experimental workflow has been introduced to mimic the rapid solidification of AM by laser scanning cast/induction melted alloys [13]. If the laser-scanned cast samples could be representative of the 3D-printed samples, compositions can be selected in an expedited and inexpensive way without powder production. We thus effectively move the bottleneck step out of design iteration (see graphical abstract), i.e., the powder is only ordered if the composition is validated after several iterations on laser-scanned samples.

In this paper, we studied how much laser-scanned samples are representative of AM samples in a model alloy, presenting their differences and similarities. But firstly, we want to acknowledge that this study will not go deep into the laser-microstructure interactions within AM. Such a broad assertion requires further investigation, including an examination of thermal history, melt pool geometries, and their interactions under varying laser parameters on different alloys. Extensive research has already been conducted on single-track laser scans, with and without powder, to discern similarities and discrepancies in thermal history and melt pool behavior on different alloys [23], [24], [25], [26], [27], [28], [29]. It is expected that these problems could be more complex for multi-path laser scans. In this study, we focus on two sets of laser parameters to study if laser-scanned samples are representative of the 3D-printed ones. Special emphasis is placed on understanding to what extent the microstructures and strength of laser-scanned samples resemble the AMed ones. This is explored in great depth with the aid of multi-scale scanning electron microscopy (SEM), transmission electron microscopy (TEM), and scanning-transmission electron microscopy (STEM) characterizations. We expect this study to demonstrate the opportunities and limitations of the expedited experimental workflow for a model Al-Ni-Zr-Er alloy system while encouraging a broader investigation into the various laser parameters and different alloy systems in the following studies.

## 2. Material system

In this study, an Al-2.73Ni-3.19Zr-2.34Er wt% model alloy, designed in our earlier study is used [13]. The alloy was designed for high-temperature (250°C) applications. It showcases a great coarsening resistance demonstrating stable yield strength and microstructure after aging 48 hours at an elevated temperature of 400°C. The yield strength is 400 MPa at a peak aging time of 8 hours at 400°C. ThermoCalc software was employed to simulate the material system. Specifically, Scheil calculation was conducted to mimic the rapid solidification process and simulate the as-built system, while a single equilibrium calculation was performed to simulate the material system under service conditions at 250°C. The alloy system consists of various phases, including Al-FCC, $Al_3Ni$, $Al_3M$ (in $L1_2$ structure and $D0_{23}$ structure), and $Al_{23}Ni_6M_4$, where M is either Er or Zr. It is noteworthy that the ternary $Al_{23}Ni_6M_4$ phase diminishes to 0 at the equilibrium state because it is a metastable phase that can transform into $L1_2$ (initially absent in the as-built system) and $Al_3Ni$ phases after an extended aging period. The Scheil solidification curve (Fig. S1) and phase fraction at as-built and in-service states (Table. S1) are shown in *Supplementary* section.

This material system is primarily governed by two prominent strengthening mechanisms: grain boundary strengthening mechanism (Hall-Petch effect, Eq. 1) and precipitation strengthening mechanisms (Orowan strengthening, Eq. 2). In Hall-Petch effect, Al grain boundaries act as impediments to the motion of dislocations, thus contributing to the augmentation of material strength, a phenomenon described by Eq. 1:

$$\sigma_{HP} = \sigma_0 + \frac{K}{\sqrt{d}} \qquad 1$$

$\sigma_0$ is the material constant for the frictional stress in Al lattice, *d* is the grain size, and *K* is the Hall-Petch coefficient depending on material systems. On the other hand, Orowan strengthening primarily stems from precipitate phases. When external forces are applied, these particles interact with dislocations, impeding their sliding and obstructing their forward advancement, thus, strengthening the material. The increased strength achieved through the mechanism of Orowan dislocation looping around these non-shearable precipitates is as shown in Eq. 2 [30]:

$$\Delta\sigma_{Or} = Q \frac{0.4Gb}{\pi\sqrt{1-v}} \cdot \frac{\ln(\frac{2\bar{R}}{b})}{\bar{\lambda}} \qquad 2$$

where $Q$ is the Taylor factor for Al matrix, $G$ is the shear modulus, $b$ is the magnitude of Burgers vector, $v$ is the Poisson ratio, $\bar{R} = \sqrt{\frac{2}{3}} <r>$ is the mean planar precipitate radius ($<r>$ is the precipitate mean radius), and $\bar{\lambda} = (\sqrt{\frac{3\pi}{4f}} - 1.64)\bar{R}$ is the mean planar distance between precipitates, in which we assume a homogenous distribution of spherical precipitates on a cubic grid [31]. Here $f$ is the volume fraction of the precipitates. As the volume fraction of precipitates increases and their size decreases, the contribution of Orowan strengthening to the material system intensifies. However, there is a critical threshold to consider, where precipitates that are too small, particularly those below 2 nm in size [13], may allow dislocations to shear through them, thereby diminishing the overall strengthening effect. Therefore, the objective was to achieve precipitate sizes that are as small as possible while remaining above the 2 nm threshold. This balance ensures an optimal combination of strengthening mechanisms, maximizing material performance while avoiding the detrimental effects of excessively small precipitates.

## 3. Experiment Method

### 3.1 3D-printed sample preparation

The powder was generated by AMAZEMENT company through ultrasonic atomization using the AUS500 system of Indutherm Bluepower. The particle size was characterized using the laser diffraction method, revealing a distribution with d10 = 44.5%, d50 = 63.8%, and d90 = 91.5%. The chemical composition of the powder was confirmed by X-ray fluorescence analysis.

Cubic samples measuring 6 mm in length were additively manufactured utilizing an SLM Solutions 250 HL machine. This system features a solid-state Nd: YAG laser operating at a wavelength of 1064 nm, with a maximum laser power of 400 W. The focus beam size is 70 μm with a Gaussian distribution. Before printing, the powder underwent vacuum drying to reduce the relative humidity to below 5%. The chamber was filled with Ar-4.6 to maintain an oxygen level below 1000 ppm to prevent oxidation. The build plate was heated to 200°C to facilitate adhesion. The processing parameters were optimized to minimize printing defects, including a laser power of 350 W, a hatch distance of 120 μm, a scan speed of 1100 mm/s, a layer thickness of 50 μm, a scan strategy involving 8 mm stripes, and a rotation angle of 67°. These parameters are summarized in Table 2 and were compared with those used for the laser-scanned samples.

### 3.1 Laser-scanned induction-melted sample preparation

The alloy with Al-0.4Er-1Zr-1.33Ni at. % was melted employing the MC20V induction heating system from Indutherm. Granulated elements with diameters less than 5 mm were utilized, with each batch comprising 30 g of raw elements. Prior to melting, a boron nitride (BN) spray was utilized to coat the surface of the $Al_2O_3$ crucible, serving to minimize the interaction between the melting elements and the crucible. A vacuum environment of 0.1 mbar was applied to reduce oxygen levels, and to prevent oxidation, argon gas was introduced to fill the chamber. The temperature was then elevated to 1000°C to ensure complete material melting, which was maintained for 5 minutes. Subsequently, the molten material was allowed to naturally cool inside the chamber after the heat source was turned off, shaping it into rod structures with approximately 60 mm in length and 15 mm in diameter.

These rods were further processed by rolling them into metal sheets through a series of 10 iterations, gradually reducing the rolling gap until achieving a thickness of 2 mm. During rolling, the samples were heated to 100°C to prevent edge cracking. Following rolling, a subsequent pressing operation was conducted to flatten the slightly bent sheets.

Subsequently, the metal sheets underwent laser scanning with two sets of laser parameters: The first was scanned by using the LT30 LPBF system from DMG MORI. The system is equipped with a solid-state Nd: YAG laser source emitting at a wavelength of 1064 nm, with a maximum laser power of 1000 W. The focus laser beam has a diameter of 70 μm with a Gaussian distribution. The chamber was maintained under vacuum and flowed with argon gas at 1 atm pressure, with a purity of 4.6, ensuring oxygen content inside the chamber remained below 1000 ppm. The processing parameters are scan speed of 300 mm/s, hatch distance of 170 μm, and laser power of 400 W. The first parameters were chosen to maximize both the density and the melt pool depth for the convenience of subsequent characterizations. The second sample was scanned by EOS M290 system. The system is equipped with Yb-fiber laser emitting at a wavelength of 1064 nm with a maximum powder of 400 W. The beam diameter is 100 μm. Inert argon atmosphere was maintained in the build chamber, ensuring a maximum oxygen concentration of 0.10%. The processing parameters are a laser power of 350 W, a hatch distance of 120 μm, and a scan speed of 1100 mm/s. The second laser-scanned sample is a controlled sample having the same parameters as the 3D-printed sample, although having a shallow melt pool constraints its ability to comparison in subsequent characterizations. The workflow of the laser-scanned samples preparation is depicted in Fig. 1, and laser parameters, including the 3D-printed one, are listed in Table 1.

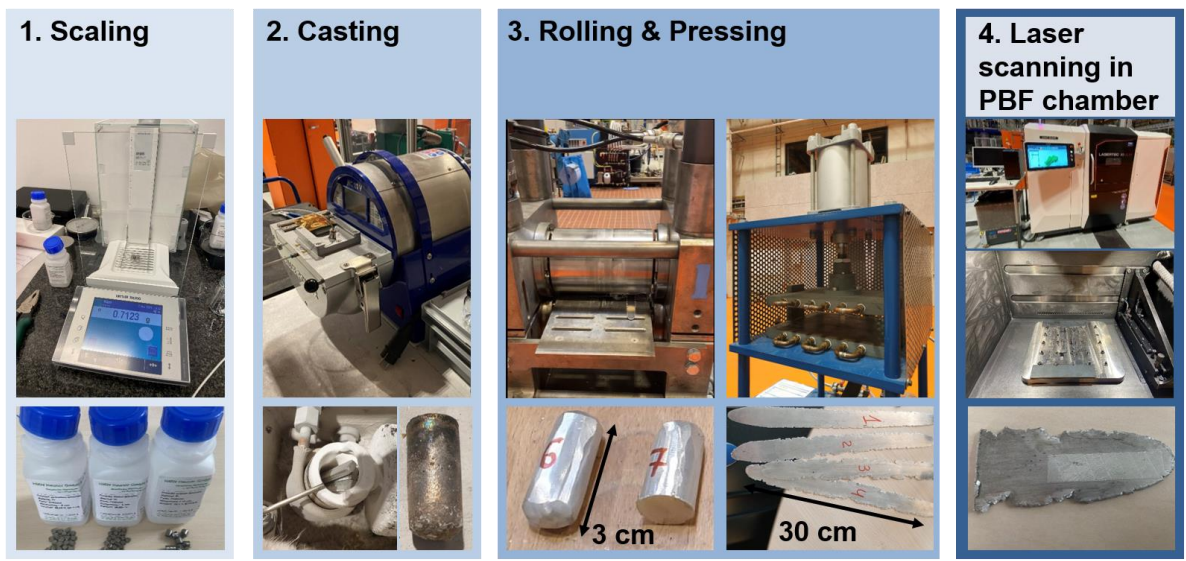

*Fig. 1* *Workflow for laser-scanned sample preparation. Above pictures and laser-scanned sample 1 were prepared by Florian Hengsbach.*

*Table. 1* *Processing parameters of the 3D-printed and laser-scanned samples.*

|  | **3D-printed sample** | **Laser-scanned sample 1 (LS#1)** | **Laser-scanned sample 2 (LS#2)** |
|---|---|---|---|
| Scan speed | 1100 mm/s | 300 mm/s | 1100 mm/s |
| Hatch distance | 120 μm | 170 μm | 120 μm |
| Laser power | 350 W | 400 W | 350 W |
| Powder thickness | 50 μm | NA | NA |
| Preheat | 200°C | Room temperature | Room temperature |
| Note* | Maximizing printing density | Maximizing melt pool depth to help subsequent characterizations, meanwhile ensure maximum density | Control parameters which are the same as 3D-printed sample |

### 3.3 Etching

To characterize melt pools, samples were polished and then etched by Keller etchant (2 mL HF, 3 mL HCl, 5 mL $HNO_3$, and 190 mL $H_2O$) for 15 s. The samples were immediately rinsed by water for 30 s.

### 3.4 SEM characterization

SEM images, SEM-EDS, and EBSD data were acquired by ThermoFisher Apreo HiVac machine, equipped with an EDAX Elite 150 SDD EDS detector and EDAX Hikari EBSD detector.

Both EDS and EBSD data were acquired at 15 kV and 3.2 nA. The EDS spectra and maps were processed by EDAX APEX software. EBSD data were firstly re-indexed by Dictionary Index to improve indexing fidelity [32], and then were performed grain selection using MTEX package on MATLAB.

### 3.5 TEM sample preparation and characterization

Laser-scanned samples were prepared by mechanical polishing. Initially, the samples were polished using the Allied MultiPrep Polishing System to achieve a thickness of approximately 50 µm. Olympus SZX12 optical microscopy was employed to check the thickness. Subsequently, Ar-ion milling was carried out using the PIPS II system. The ion gun energy was set to 10 kV until pierced holes became visible. Thereafter, the energy was progressively reduced from 2 to 0.5 kV, until the desired thickness was obtained. Low-energy beam cleaning to reduce ion-induced damage was performed at an angle of 2° for a duration of 10 minutes for each energy. STEM images were acquired using a Thermo Fisher Themis 200 G3 instrument equipped with a probe aberration corrector. Images were captured at an accelerating voltage of 200 kV with a probe convergence semi-angle of 17.9 mrad. HAADF images were acquired using 70-200 mrads collection angles and drift-corrected using cross-correlation to reduce signal-to-noise ratio. EDS was acquired using a Thermo Fisher SuperXG2 detector, and the resulting data was processed using Thermo Fisher Scientific Velox software V3.6.0. Low magnification EDS data was collected with a probe current of 140 pA. A set of 211 frames were collected and drift-corrected using cross correlation. The data was post-filtered using the Gaussian method. Atomic-resolution EDS mapping was collected with a screen current of 130 pA. A set of 1465 frames was collected and drift-corrected using cross correlation.

For the 3D-printed sample analysis, the specimen was initially extracted using a Helios Nanolab 660 focused ion beam microscope manufactured by Thermo Fisher Scientific. A lamella of the targeted region was then isolated and affixed onto a molybdenum TEM grid using an Omniprobe 400 micromanipulator from Oxford Instruments. Subsequently, the sample was thinned by gradually reducing the ion beam voltage from 30 kV to 16 kV until the thickness reached 300 nm. The beam voltage was then halved until the thickness was further reduced to 150 nm. Finally, the sample underwent final polishing at 750 V until reaching a thickness of approximately 50 nm, utilizing a Model 1040 NanoMill from Fischione Instruments. The ion damage was cleaned with Ar-ion milling by Fischione 1051 TEM Mill from Fischione Instruments. The cleaning process involved using beam voltages of 0.3 and 0.1 kV for durations of 3 and 1 minute, respectively. Characterization of the sample was conducted using the Themis Z probe aberration-corrected TEM/STEM system from Thermo Fisher Scientific. STEM images were acquired at a voltage of 300 kV, a beam current of 40 pA, and a probe convergence semi-angle of 18.8 mrad. High-angle annular dark-field (HADDF) images were obtained using a collection semi-angle of 78-200 mrad with drift-correction implemented via the Revolving STEM method [33]. Energy-dispersive EDS analysis was performed using a Thermo Fisher SuperX detector, and the resulting data was processed using Thermo Fisher Scientific Velox software. Low-magnification EDS data was acquired using a beam current of 200 pA and processed using a 5-pixel averaging filter. Atomic EDS data acquisition employed a beam current of 50 pA and was filtered using non-local principal component analysis [34].

### 3.6 Hardness test

Microhardness of the laser-scanned sample and 3D-printed sample were measured by Wilson VH3100 from Beuhler. For both samples, 10 measurements were performed under 100 g loading force, and the average hardness was taken. The indentation depth was calculated to ensure that the penetration is within the rapid solidification features, e.g., melt pool depth, of the laser-scanned sample. The indent of Vickers hardness is a standard 136° pyramidal diamond. The penetration depth was calculated from the square indent based on geometry as described by Eq. 3,

$$t = \frac{d/2}{\tan(136°/2)} = 0.202 \cdot d \qquad 3$$

where $t$ is the penetration depth and $d$ is the diagonal of the square indent.

## 4. Results and discussion

The optical images of the melt pool geometries of three samples—3D-printed, LS#1, and LS#2—are presented in Fig. 1. Each layer in the 3D-printed sample experienced a complex thermal history, as successive layers of powder were subjected to multiple heating cycles at varying angles due to the rotation of the laser beam. Moreover, local distribution of powder itself may add to the stochasticity of the melt pool, reflected by more variations in melt pool geometries as shown in Table. 2. In contrast, the bulk laser-scanned samples underwent a multiple organized laser pass in only one layer, resulting in a more uniform and linear arrangement of melt pools compared to the disorganized melt pool structure observed in the 3D-printed sample. The geometrical attributes of the melt pools, including melt pool width, depth, and aspect ratio (width-to-depth ratio, W/D), are summarized in Table 2. LS#1 sample, characterized by higher energy density due to slower scan speed and higher laser power, exhibited a keyhole mode melt pool geometry with a narrow and deep profile, as shown in Fig. 2(b). LS#2 and 3D-printed samples share the same laser power, scan speed, and hatch distance, exhibiting similar melt pool geometries. However, LS#2 displays larger melt pool dimensions—greater width and depth—compared to the 3D-printed sample.

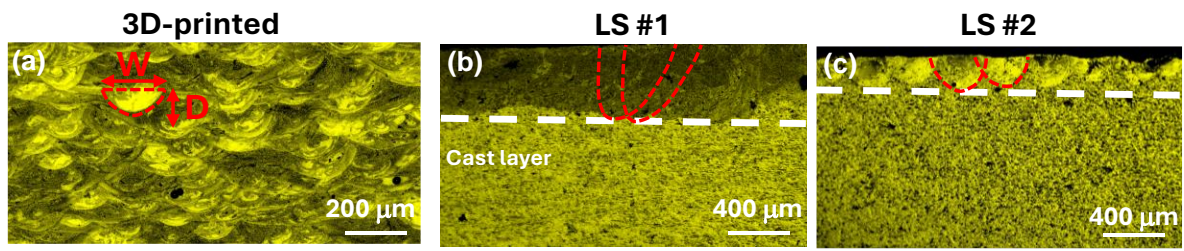

*Fig. 2 Melt pool characterizations of the three samples by optical microscope. LS#1 has narrow and deep melt pool profile indicating keyhole mode due to large energy input. LS#2 has the same laser parameters with the 3D-printed, thus exhibiting more similar melt pool shapes.*

The discrepancy could be derived from the powder-driven effect, such as entrainment which can affect melt pool behaviors [35], [36], [37]. As the interpass temperature builds up due to thermal history, the metal vaporization rate and powder flowability increases, which subsequently enhances the powder-entrained gas flow and changes melt pool geometries [38], [39]. Another plausible explanation is the difference in beam diameter between the systems used: the

LS#2 sample was processed using the EOS M290 system with a beam diameter of 100 µm, whereas the 3D-printed sample was processed using the AUS500 Indutherm Bluepower system with a beam diameter of 70 µm. The larger beam diameter in the LS#2 sample could account for the wider melt pool.

*Table. 2 Melt pool dimensions in the three samples*

|  | Width, $W$ | | Depth, $D$ | | Aspect ratio, $\frac{W}{D}$ | |
|---|---|---|---|---|---|---|
|  | Average (µm) | Variation | Average (µm) | Variation | Average (µm) | Variation |
| **3D-printed sample** | 161.9 | 12.8 % | 59.8 | 20.7 % | 2.77 | 11.9 % |
| **Laser-scanned sample 1 (LS#1)** | 319.2 | 2.27 % | 480.3 | 9.06 % | 0.66 | 10.67 % |
| **Laser-scanned sample 2 (LS#2)** | 335.4 | 9.30 % | 179.1 | 14.65 % | 1.87 | 11.6% |

At the SEM scale, the microstructural morphology appeared to be similar across the three samples (Fig. 3). Melt pool boundaries were delineated with dashed lines, and the microstructures within the melt pool and at the melt pool boundaries or fusion boundaries were highlighted with blue and red boxes, respectively. In all three samples, the melt pool interiors were predominantly characterized by cellular structures, whereas the melt pool boundary exhibited fine, equiaxed grains. This phenomenon aligns with observations in welding studies on steels [40]. The behavior can be attributed to the thermal gradient distribution during solidification. High thermal gradients and high cooling rate near the heat source (within the melt pool) promote the formation of cellular microstructures, whereas lower thermal gradients farther from the heat source (at the melt pool boundaries) favor equiaxed grain growth [41]. However, notable differences were observed between the laser-scanned and 3D-printed samples. Large precipitate islands, present in the laser-scanned samples, were absent in the 3D-printed sample. These precipitates likely formed during induction melting under slower cooling rates compared to LPBF processing and were too large to be fully remelted by the laser beam even for the high energy in LS#1, thereby retaining their original morphology. EDS analysis (*Supplementary* Fig. S2) identified these precipitates as primarily $Al_3Zr$ and ternary phases. In contrast, during powder-based 3D printing, the smaller size of powder particles and high cooling rates allow for much smaller precipitates. To minimize macrosegregation of precipitate phases, we employed induction melting to cast 30 g per batch. The whirling dynamics generated by the wire inductor effectively mitigate macrosegregation in such small sample sizes. Although post-heat treatments can further address this issue, newly designed alloys require step-specific optimization, which would conflict with the objectives of an expedited design process. Nonetheless, we encourage readers to consider applying post-heat treatments to laser-scanned samples to achieve an even closer representation of AM samples, acknowledging that this approach may come at the expense of experimental efficiency.

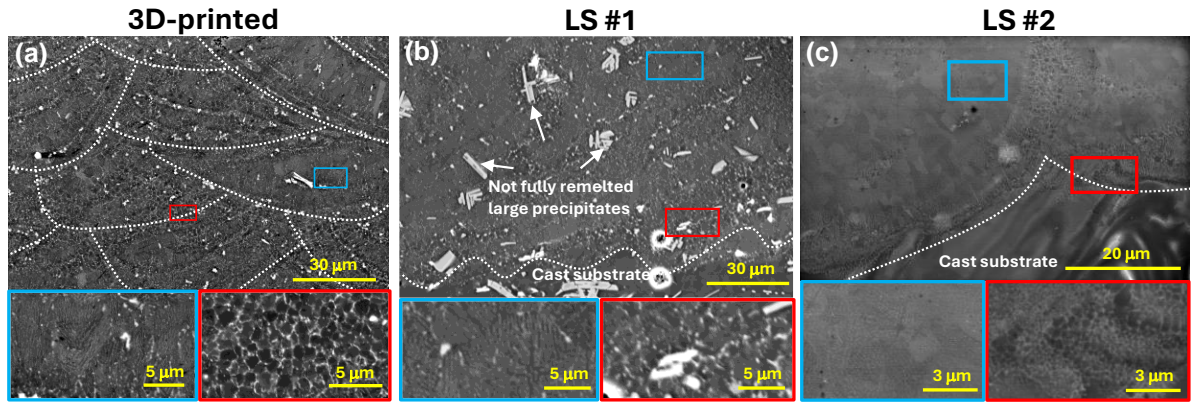

***Fig. 3*** *SEM images of the three samples. Melt pool boundaries are delineated with white dashed lines. Microstructures within the melt pool interior and at the melt pool boundaries are highlighted with blue and red boxes, respectively. Melt pool interiors are dominated by cellular microstructures due to high thermal gradient and high cooling rate, whereas the melt pool boundaries and HAZ with low thermal gradient are dominated by equiaxed grains.*

EBSD analysis was conducted to examine the texture and grain size distribution of the samples (Fig. 3). In the 3D-printed sample (Fig. 3(a)), elongated grains were observed, extending from near melt pool boundaries into the melt pool interior due to epitaxial growth. Fine equiaxed grains were predominantly located at the melt pool boundaries. For the laser-scanned samples (Figs. 3(b) and 3(c)), a clear interface between the laser-scanned region and the cast-and-rolled region was observed and marked by dashed lines. Pole figures in the <100> direction was plotted, with blue-bordered figures representing orientations in the laser-scanned regions and red-bordered figures corresponding to orientations in the as-rolled regions. It is seen that the laser remelting process in the laser-scanned samples resulted in smaller grains and randomized grain orientations, even though the substrate retained a strong rolling texture. However, unlike the 3D-printed sample, elongated grains formed by epitaxial growth were not observed in the laser-scanned samples, potentially due to a specific textured of the substrate. Moreover, less fraction of equiaxed fine grains were present. This difference can be attributed to the distinct thermal histories of the samples. The 3D-printed sample experienced multiple heating cycles, promoting extensive recrystallization in fusion boundary, resulting in a higher fraction of equiaxed fine grains compared to the laser-scanned samples. We also compared the grain size distribution across the samples. Notably, the grain size distribution was relatively similar (Fig. 5), with most grains falling within the range of 1–2 μm. Although laser-scanned samples exhibited slightly larger average grain sizes compared to the 3D-printed sample, they remained within a comparable range. LS#2, processed using the same laser parameters as the 3D-printed sample, had an average grain size of 1.34 μm, representing an 11.7% difference from the 3D-printed sample's average of 1.20 μm. This is expected that the LS#2 sample has a more similar grain size to 3D-printed sample in comparison to LS#1 sample. LS#1, despite being processed with significantly different laser parameters, displayed an average grain size of 1.52 μm which has a 26.7% difference relative to the 3D-printed sample. The average is not significantly different and can be attributed to the high power and slow speed [42], [43].

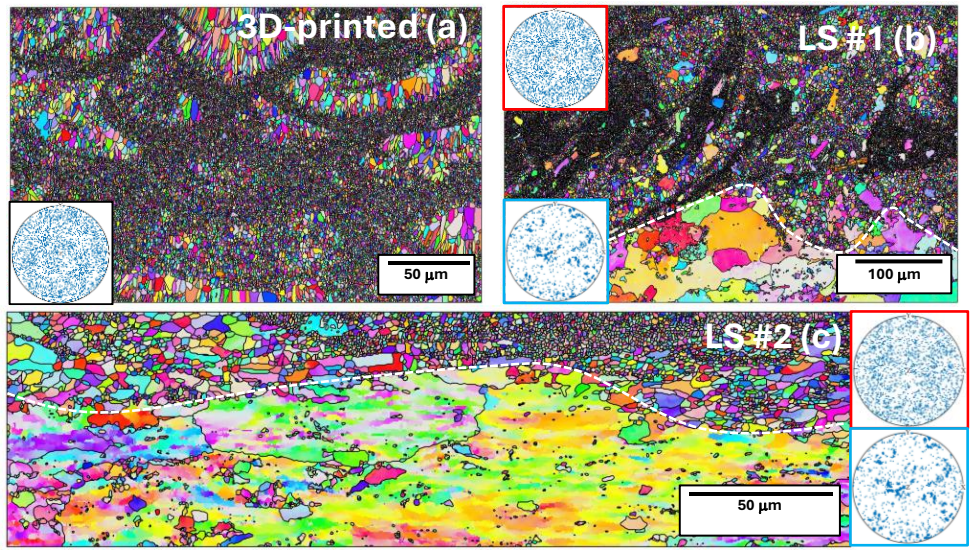

***Fig. 4*** *EBSD inverse pole figure (IPF) maps of **(a)** the 3D-printed sample, **(b)** LS#1, and **(c)** LS#2. Insets show <100> pole figures. For the laser-scanned samples, pole figures with red borders represent orientations in the laser-scanned regions, while those with blue borders correspond to orientations in the rolled cast substrate. The laser-scanning process is observed to remove the textured microstructure, effectively "shuffling" the grain orientations and resulting in significantly smaller grain sizes.*

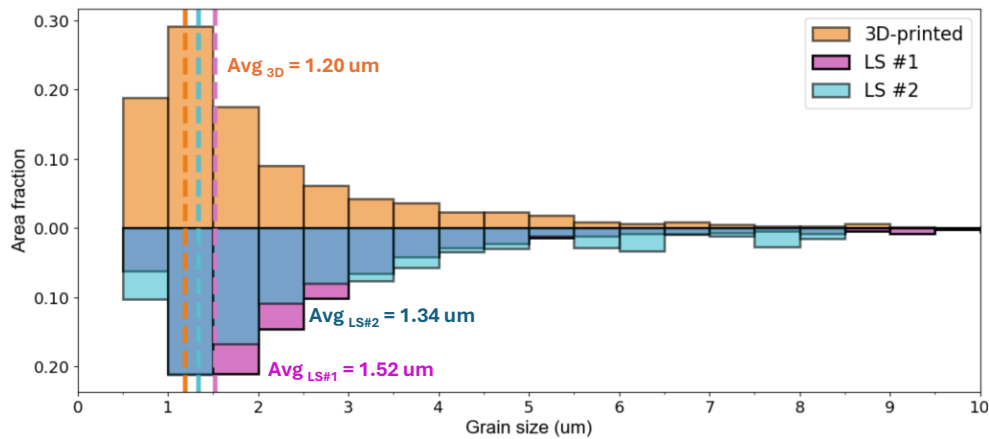

***Fig. 5*** *Grain size distribution of the 3D-printed sample, LS#1, and LS#2. All samples exhibit a similar distribution, peaking in the range of 1–2 µm. The laser-scanned samples show slightly larger grain sizes compared to the 3D-printed sample. LS#2, processed with the same laser parameters as the 3D-printed sample, shows closer average grain size values. Despite the significant difference in laser parameters, LS#1 also does not demonstrate significant change in average grain size to the 3D-printed sample.*

We then conducted SEM-EDS analysis at high magnification on the regions with fine and equiaxed grains (Fig. 6), where there is a relatively high degree of element micro-segregation at grain boundaries. At this scale, we can observe the presence of Ni-rich ($Al_3Ni$) and Zr-rich ($Al_3Zr$) precipitates in the specimens. Additionally, segregation of Ni and Er along grain boundaries was observed. These regions rich in Ni and Er are anticipated to correspond to $Al_{23}Ni_6M_4$ ternary precipitates predicted by CALPHAD simulations. Notably, this ternary phase is a metastable phase which is only associated with rapid solidification processes.

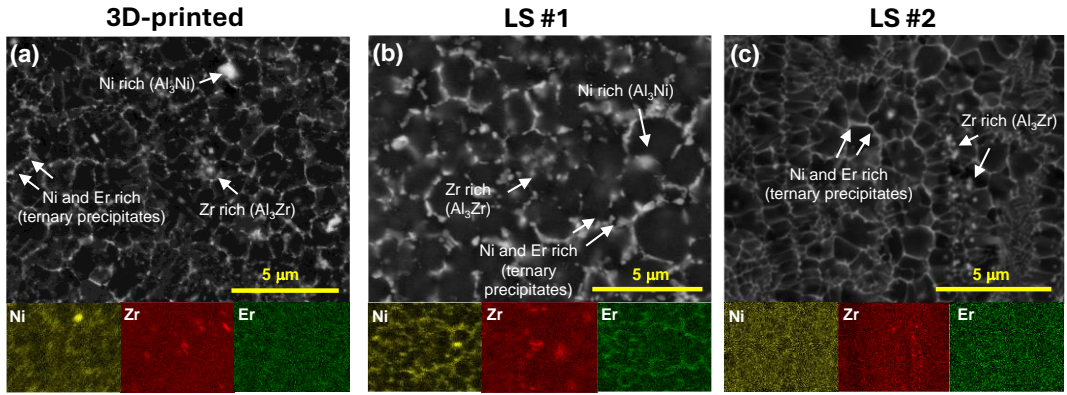

*Fig. 6* SEM-EDS images of *(a).* 3D-printed sample and *(b-c).* laser-scanned samples. $Al_3Ni$, and Al-Ni-Er ternary phases were observed in both 3D-pritned and laser-scanned samples.

To compare the local microstructure, STEM imaging and STEM-EDS analysis were performed on a single grain from each sample (Fig. 7). Due to the shallow melt pool in the LS#2 sample, TEM specimens for the laser-scanned samples were prepared only from the LS#1 sample, which was used for subsequent TEM comparisons with the 3D-printed sample. At this grain-scale magnification, the microstructure and phase distributions of the two specimens appeared remarkably similar. Specifically, Al-Ni-Er ternary phases were observed to be distributed along the grain boundaries in both samples, consistent with the observations made through SEM-EDS analysis.

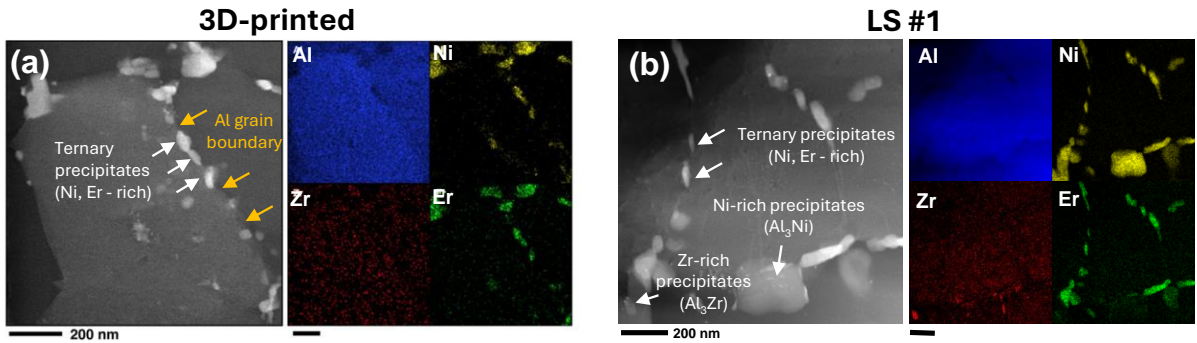

*Fig. 7* Low-magnification STEM-EDS images of *(a).* 3D-printed sample and *(b).* LS#1. The Al-Ni-Er rich ternary phase is located on grain boundaries in both samples. The images of 3D-printed sample are from [13].

Upon closer examination of the Al grain interiors, $L1_2$ nanoprecipitates were observed by atomic STEM in [110] zone axis in both 3D-printed sample and LS#1 sample (Fig. 8). Coherent interfaces were observed between the $L1_2$ nanoprecipitates and the Al matrix in both cases. Additionally, a core-shell-like structure was identified within the $L1_2$ nanoprecipitates in both samples. This structure revealed a higher concentration of Zr in the shell, while Er was more concentrated in the core. This core-shell structure agrees to existing literature [44], [45]. The formation of the Zr/Er segregation in the core-shell structure was previously justified through first principal density functional theory (DFT) by Zhang et al [46]. The formation of the core-shell structure can also be elucidated by considering the lattice parameter of Zr exhibiting more similarity to that of the Al matrix, which makes it energetically favorable to the distribution in contact with the Al matrix. In this structure, the slower diffusion rate of Zr within the Al matrix at the shell effectively encapsulates the more rapidly diffusing Er within the core. This configuration enhances coarsening resistance, which explains the robustness of the material's strength even at temperatures up to 400°C [47], [48], [49].

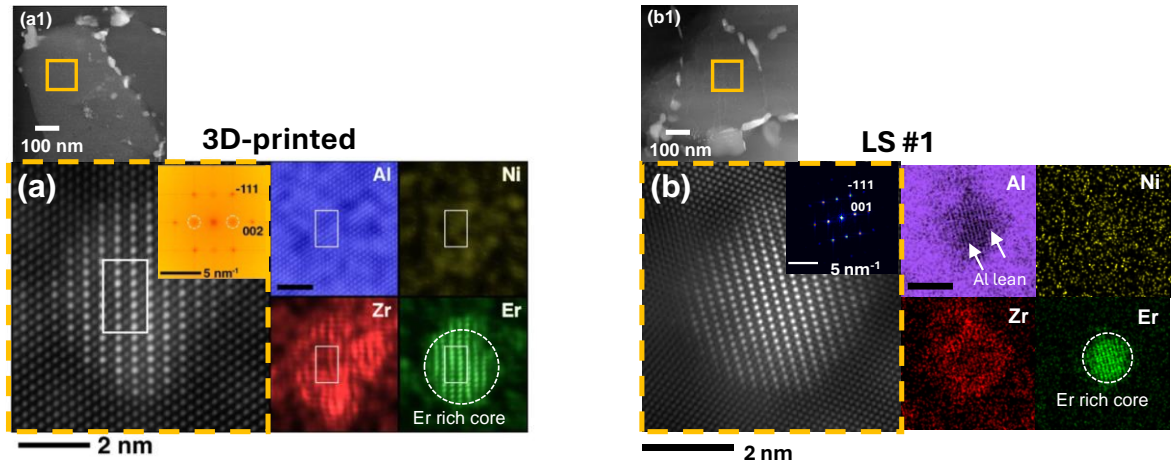

*Fig. 8* Atomic STEM-EDS images of *(a).* 3D-printed sample and *(b).* LS#1. Both samples show the presence of L1$_2$ nanoprecipitates. Besides, the precipitates in both samples exhibit a core-shell configuration with Zr rich at the core and Er rich at the shell. The images of 3D-printed sample are from [13].

The sizes of ternary phase and L1$_2$ nanoprecipitates were measured and were presented in Table 3. Despite the use of different laser parameters to produce the samples, LS#1 exhibits a precipitate size comparable to that of the 3D-printed sample. Notably, the L1$_2$ nanoprecipitates exhibit only 4% sizes difference across both samples, while the dimensions of ternary precipitates in the laser-scanned sample exhibits a respective increase of 14% compared to those in the 3D-printed.

*Table. 3 Phase size comparison between the two 3D-printed sample and LS#1*

|  | Ternary precipitates (nm) | L1$_2$ nanoprecipitates (nm) |
|---|---|---|
| **3D-printed sample** | 45.1 ± 6.7 | 2.5 ± 0.7 |
| **Laser-scanned sample** | 51.6 ± 16.4 | 2.4 ± 0.7 |
| **Difference** | 14.4 % | 4 % |

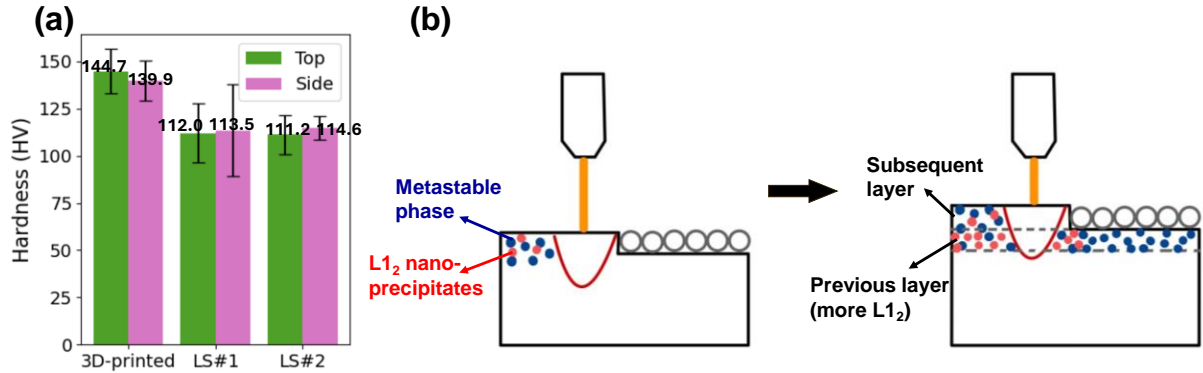

*Fig. 9 (a).* Vickers hardness of the 3D-printed sample and laser-scanned samples from the top (parallel to the building direction) and side (perpendicular to the building direction). No significant anisotropy was observed. Laser-scanned samples had an overall ~20% decrease in hardness compared with the 3D-printed sample. *(b).* Schematic of metastable phase transformation during LPBF. The thermal history parallel to building direction facilitate the transformation of metastable phase formed in the previous layer into $L1_2$ nanoprecipitates which enhance the material hardness.

The hardness of the samples was measured to evaluate their mechanical performance. Measurements were taken in two orientations: parallel and perpendicular to the building direction, to assess anisotropy and determine if the laser-scanned samples replicate the behavior of the 3D-printed sample. For the tests parallel to the building direction, we followed the general indentation rule for thin film samples that the indentation depth must be within 10% of film thickness to reduce the substrate effect [50], [51]. In this case, laser-scanned regions are analogous to thin-film which has fine microstructure and different mechanical properties to the unscanned region which can be analogous to the substrate. Besides, to avoid erasing laser-scanned features from the top (parallel to the building direction), the hardness was measured on the as-scanned surface without polishing. The average indent diagonal in the laser-scanned samples was 41 μm, corresponding to an average indent depth of 8.3 μm based on Eq. 3. This value is within 10% of the melt pool depth for both laser-scanned samples (480 μm for LS#1 and 179 μm for LS#2). The hardness results are presented in Fig. 9(a). The 3D-printed sample exhibited a hardness of 144.7 HV along the top surface and 139.9 HV along the side surface (perpendicular to the building direction), indicating no significant anisotropy. This lack of anisotropy was also observed in the laser-scanned samples, with LS#1 measuring 112.0 HV (top) and 113.5 HV (side), and LS#2 measuring 111.2 HV (top) and 114.6 HV (side). It is noteworthy that the hardness of LS#1 and LS#2 were nearly identical, despite being processed with significantly different laser parameters. However, a notable difference was observed between the laser-scanned samples and the 3D-printed sample in both directions. The average hardness of the 3D-printed sample was 142.3 HV (calculated as the mean of both orientations), whereas the average hardness of the laser-scanned samples was 112.8 HV (averaged across both samples and orientations), reflecting a 20.7% reduction compared to the 3D-printed sample. Two potential factors may account for this discrepancy:

1). The phase fraction of the strengthening phases could be different. According to Eq. 3, Orowan strengthening depends on both precipitates size and phase fractions. However, in this material system, the precipitates phase fraction is very difficult to measure. The crystallography information of the ternary $Al_{23}Ni_6Er_4$ metastable phase is still undocumented in literature and material databases. More TEM work needs to be done to characterize its crystal structure. Besides, the precipitates phase size is so small in nanometer size. All these factors make it hard for laboratory X-ray diffraction as a general approach to phase fraction characterization.

2). In LPBF printing, the cyclical thermal exposure along the build direction induces a transformation from the metastable ternary phase to $L1_2$ phase. Fig. 9 (b) presents the schematic, while the experimental observations by TEM can be referred to [13]. The $L1_2$ phase, characterized by its nanometer size, enhances material strength through the Orowan mechanism (Eq. 2). Consequently, the increased presence of the $L1_2$ phase accounts for the elevated hardness observed in the 3D-printed sample.

One might argue that the difference in Al grain size contributes to the hardness variation between the 3D-printed sample and the laser-scanned samples. However, it is essential to note that the two laser-scanned samples, despite having different grain sizes, exhibit nearly identical hardness values. This observation suggests that the grain size differences are too small to significantly influence hardness compared to other contributing factors.

To validate our methodology, we also compared the hardness of the laser-scanned and 3D-printed sample of a printable benchmark alloy (Al-Ni-Zr-Er-Y-Yb) [11]. The hardness of the 3D-printed sample was obtained from reference [11]. The laser-scanned sample used for comparison was prepared under identical conditions to LS#1, including the same casting, rolling procedures, and laser parameters. The hardness measurement results are presented in Fig. 10. It is worth noting that the reference did not specify the measurement direction for the hardness values. Thus, we averaged the hardness measurements from both directions for the laser-scanned samples of both the benchmark alloy and the model alloy. We found that the hardness variance due to different processing in our designed alloy was also observed in the benchmark alloy. In our designed alloy, the laser-scanned sample demonstrated a 20.7% decrease in hardness relative to its 3D-printed counterpart. A comparable trend was observed for the benchmark alloy, with a 30.2% decline in hardness for the laser-scanned sample, demonstrating a ~10% absolute error compared with our model alloy.

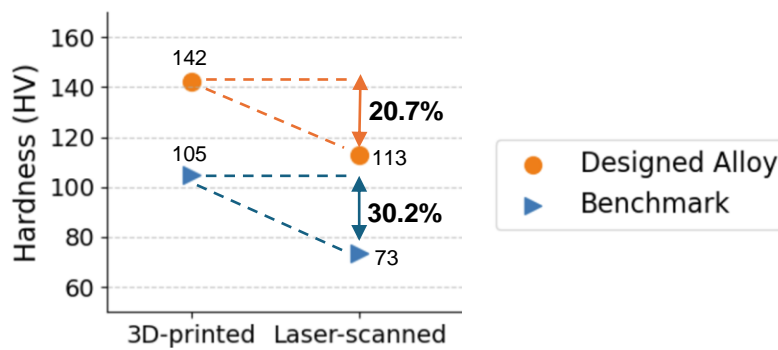

**Fig. 10** *Comparison of the microhardness between the designed alloy and the benchmark alloy for both laser-scanned and 3D-printed samples. Both alloys exhibit a similar trend, with an absolute error of approximately 10%.*

Although laser-scanned samples cannot fully replicate the hardness of LPBF samples, they demonstrate a consistent trend of 20–30% hardness reduction compared with LPBF samples, which offers valuable insights for advancing alloy design in the realm of AM by providing an economical and accelerated approach for screening compositions, circumventing customized powder production usually as the bottleneck of the workflow.

For example, in the context of designing creep-resistant or high-temperature alloys, the hardness trends observed in laser-scanned samples at varying temperatures can serve as indicative markers. If a composition exhibits a significant decrease in hardness with increasing temperatures, it is likely unsuitable for such applications and can be eliminated early in the design process. Additionally, laser-scanned samples offer insights into the microstructure of AM samples. While the melt pool geometries differ even under identical processing parameters, the microstructures—down to SEM and TEM scales—are highly similar, especially morphology, grain size, types of precipitates, and phase distributions. Nevertheless, this study has limitations. The current investigation only focused on comparing two alloy systems, with a particular emphasis on Al alloys. The essence of this study is to encourage broader exploration into different alloy systems and more laser parameters to enrich the knowledge of alloy design in AM community.

## 5. Conclusion

To address the constrained availability of customized powder in the field of AM alloy design, this paper studied Al-Ni-Zr-Er as a model alloy and investigated how much laser-scanned induction melted samples, which is a rapid experimental workflow and aiming at mimicking rapid solidification of AM, can be representative of LPBF sample. Multi-scale characterization from OM down to STEM was performed on one LPBF sample and two laser scanned samples – LS#1 (different laser parameters as LPBF sample, but maximizing both melt pool depth and density for subsequent characterization) and LS#2 (the same laser parameters as LPBF sample) to compare the microstructures of them. Hardness tests parallel and perpendicular to the building directions were conducted to evaluate mechanical performance. There are several major conclusions:

a. At an optical microscope (OM) scale (×20 magnification), laser-scanned samples exhibit larger melt pool widths and depths compared to the LPBF sample, despite identical laser parameters. Specifically, the LPBF sample has a melt pool width of $W_{LPBF} = 162$ μm and a depth of $D_{LPBF} = 60$ μm, while LS#1 has $W_{LS1} = 319$ μm and $D_{LPBF} = 480$ μm, and LS#2 has $W_{LS2} = 335$ μm and $D_{LPBF} = 179$ μm. The reasons discussed to be different thermal histories, powder/bulk material behaviors, and focus beam diameters.
b. At a low-magnification SEM scale (×1000 magnification), all three samples exhibit cellular microstructures within the melt pool and equiaxed fine grains at the melt pool boundaries. This behavior is attributed to changes in the thermal gradient across the melt pool. However, notable differences were observed: in the laser-scanned samples, precipitate islands inherited from the cast sample were too large to be fully melted and retained in the original shape.
c. Laser scanning removes the textured orientation induced by the rolling process and facilitates recrystallization into much finer grains, effectively mimicking the microstructure of LPBF samples. Grain size distributions, obtained from EBSD data, show a close resemblance among the samples. The LPBF sample has an average grain size of $d_{LPBF} = 1.20$ μm, LS#1 has $d_{LS1} = 1.52$ μm, and LS#2 has $d_{LS2} = 1.34$ μm , which matches more with the LPBF sample due to similar laser parameters. However, a key

d.  difference is the absence of elongated grains in the laser-scanned samples potentially due to textured substrate, a feature more present in the LPBF sample.
   d.  At high-magnification SEM scale (×12000 magnification), Ni-rich, Zr-rich, and Al-Ni-Er-enriched ternary precipitates were observed in both the LPBF and laser-scanned samples. The microstructures of the two sample types appeared highly similar at this scale.
   e.  At TEM-STEM scale (×80000 magnification), the microstructures of the LPBF and laser-scanned samples were nearly identical. Ternary precipitates were distributed along the grain boundaries, and $L1_2$ nanoprecipitates exhibited a core-shell configuration in both samples.
   f.  In the laser-scanned sample, Al grains and ternary precipitates were, respectively, 19% and 14% larger than those in the LPBF sample, and $L1_2$ nanoprecipitate size showed only a 4% difference between the two samples.
   g.  Vickers hardness test was conducted along two directions: parallel and perpendicular to the building direction for the three samples. Hardness measurements indicated isotropic behavior in both LPBF and laser-scanned samples. However, the laser-scanned samples demonstrated a hardness approximately 20% lower than that of the 3D-printed sample. Two potential reasons were proposed to account for this difference: (1) variations in phase fractions, which are challenging to characterize rapidly with current resources, and 2). the thermal history inherent in AM processing, which promotes the transformation of metastable phases into the strengthening $L1_2$ phase.
   h.  The hardness test was repeated for the benchmark alloy, comparing its LPBF and laser-scanned samples. The laser-scanned sample exhibited a consistent decrease in hardness relative to the corresponding LPBF sample, with a 10% absolute difference, highlighting a relative trend in hardness between the LPBF sample and laser-scanned samples.

In conclusion, while differences in melt pool geometries possibly due to the powder-driven effect were observed, as magnification increased to SEM and TEM scales, it revealed a high degree of similarities in microstructures between laser-scanned samples and LPBF samples. These similarities include microstructural morphologies, grain size distributions, the presence of various precipitates, and phase distributions, even when different laser parameters were applied. Especially, the key strengthening feature of this alloy, nanoscale $L1_2$ phase, were identical even in core-shell compositions, which is intriguing. Although the scope of this study was limited to specific material systems and further experiments on other material systems are required, the observed consistency in microstructure and relative hardness trends suggests that laser-scanned samples can provide valuable insights for rapid screening and validating alloy compositions in the context of Al alloy design for powder-based AM.

**Credit authorship contribution statement:**

**Zhaoxuan Ge:** Writing – Original Draft, Data curation, Formal analysis, Validation, Investigation. **Calderon Sebastian:** Data curation, Software, Writing – Review & Editing. **S. Mohadeseh Taheri-Mousavi:** Conceptualization, Methodology, Supervision, Writing – Review & Editing, Funding acquisition.

**Acknowledgements:**

The author extends gratitude to Florian Hengsbach for the induction melting, laser scanning LS#1, and printing of the sample, to James M. LeBeau and Michael Xu for their diligent efforts in characterizing the 3D-printed sample, and to Zehua Liu for re-indexing EBSD data. Special appreciation is also extended to Shaoulou Wei for his fruitful discussions and advising on the paper manuscript. We would also like to thank AMAZEMENT company for making all the powder.

Work performed in the University of Pittsburgh Nanofabrication and Characterization Core Facility (RRID:SCR_05124) and services and instruments used in this project were graciously supported, in part, by the University of Pittsburgh.

**Funding sources:**

The support for this research was provided by the Army Research Lab under Grant No. W911NF-20-2-0175. This research (or "A portion of this research") was sponsored by the Army Research Laboratory and was accomplished under Cooperative Agreement Number W911NF-20-2-0175. The views and conclusions contained in this document are those of the authors and should not be interpreted as representing the official policies, either expressed or implied, of the Army Research Laboratory or the U.S. Government. The U.S. Government is authorized to reproduce and distribute reprints for Government purposes notwithstanding any copyright notation.

**Supplementary information:**

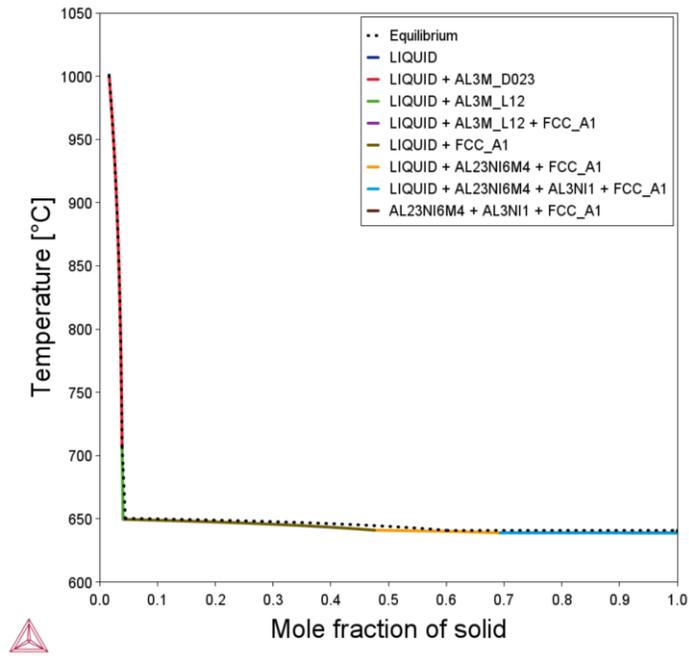

***Fig. S1*** *Phase evolution during solidification of the Al-Ni-Zr-Er material system calculated from a Scheil calculation with ThermoCalc software.*

***Table. 1*** *Stable phases and phase fraction at as-built system (Scheil calculation) and in-service system (Single equilibrium calculation) simulated by ThermoCalc.*

| Stable phases | Phase fraction (%) | |
|---|---|---|
| | As-built system | 250°C in-service system |
| $Al_3Ni$ | 2.682 | 5.32 |
| $Al_{23}Ni_6M_4$ | 3.142 | 0 |
| $Al_3M$ (D0$_{23}$) | 3.846 | 2.29 |
| $Al_3M$ (L1$_2$) | 0.173 | 3.31 |
| Al-FCC | Balance | |

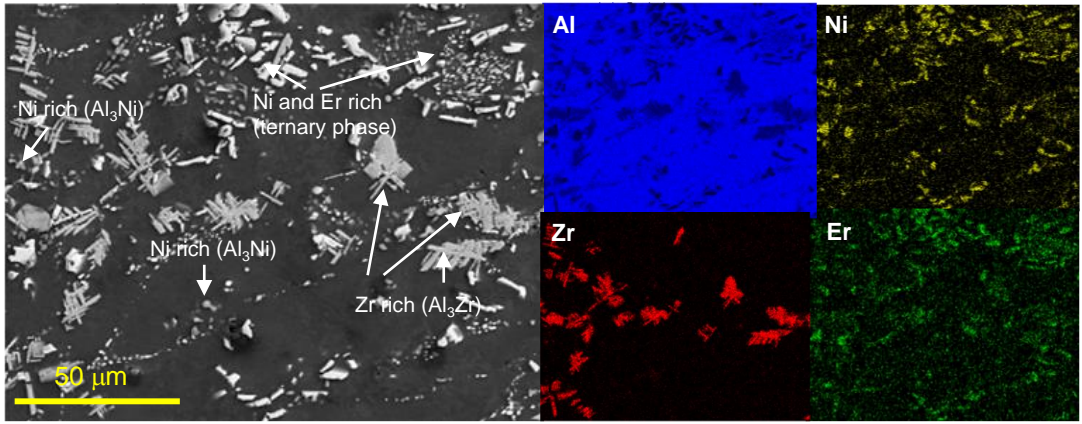

*Fig. S2* SEM-EDS images of as-rolled region. The large precipitates islands are mainly Al-Ni-Er rich ternary phase and Zr-rich phase. They are too large to be fully melted by laser beam.